\documentclass[review]{elsarticle}

\usepackage{lineno,hyperref}
\usepackage{amsmath}
\usepackage{graphicx}
\modulolinenumbers[5]

\journal{Journal of \LaTeX\ Templates}

\begin{document}

\begin{frontmatter}

\title{Stationary phase approximation for the Mach surface of superluminally moving source}
\author[VVA]{V.V.  Achkasov}
\ead{valery.achkasov@gmail.com}
\address[VVA]{Scientific Department of Fryazino Experimental Factory Ltd., Fryazino 141120, Russia}

\author[MYZ1,MYZ2]{M.Ye.  Zhuravlev\corref{cor1}}
\ead{myezhur@gmail.com}
\address[MYZ1]{St. Petersburg State University, St. Petersburg 190000, Russia}
\address[MYZ2]{Kurnakov Institute of General \& Inorganic Chemistry of RAS, 119991 Moscow, Russia}

\cortext[cor1]{Corresponding author}

\begin{abstract}
Theoretical study of superluminal sources of electromagnetic radiation boosted after the discovery of Cherenkov-Vavilov radiation. Later, the way to create fictitious sources moving superluminally was suggested. Different approaches have been proposed for the research of the distribution of the potential and the fields radiated by the superluminally moving charges. The simplest idealized cases of uniform rectilinear motion of the charge and of the charge rotating with constant angular speed open opportunities of a detailed analysis of the fields and potentials. We use Fourier series to calculate the potential distribution of point charge rotating with constant speed. An obvious advantage of this approach is that one no longer needs to calculate the retarded positions of the charge. The number of the retarded positions depends on the observation point and increases as the ratio $\omega$$R/c$ rises, where $c$ is the speed of light, $\omega$ is the rotation frequency, and $R$ is the radius of the circle. We demonstrate that equation of Mach surface can be obtained basing on the asymptotic expansion of the potential. We analyze some characteristics of the potential basing on this asymptotic expansion.  
\end{abstract}


\end{frontmatter}

\section{Introduction}
\label{S:1}

The interest in electromagnetic radiation of superluminal sources is caused by the existence of various systems in which the phase speed of a source exceeds the speed of light in a media or in vacuum. Cherenkov radiation is the most well-known among such phenomena \cite{Cher,FrankTamm} . Cherenkov radiation is an important component of different devices, e.g. loaded waveguides \cite{Jiang} or magnetooptical devices \cite{Guo}. Other systems in which the source moves with superluminal speed are the light spots moving faster than light. In particular, such is the case of pulsars’ radiation observed on Earth \cite{ArdavanApJ}. It is a common case that a detailed investigation of some phenomenon includes the study of idealized models for which an exact solution can be found, or, at least, which can be explored in details by approximate methods. The simplest idealized system in the area we study is the point charge moving rectilinearly in vacuum with a speed exceeding the speed of light. The scalar and vector potentials as well as electromagnetic field are localized inside Mach cone \cite{ITamm} which is an envelope of the emitted wave fronts. The potentials and the electromagnetic field turn into infinity on Mach cone surface. A more complicated system is a point charge rotating along a circle with constant speed. We note that the scalar potential of point charge is at the same time a Green’s function of the corresponding wave equation. Though the point charge moving with superluminal velocity does not exist \cite{ByGinzburg}, Green’s function is used to calculate the fields of the ‘fictitious’ extended superluminal sources. H. Ardavan in his series of articles \cite{ArdavanPRD,RSPA,ArdavanPRE} payed a special attention to mathematical aspects of constructing the solution of point and extended superluminal sources including point charge which rotates superluminally. Certain calculations of the field and Mach surface for the point source rotating along a circle was presented in \cite{BB}. Still, further analysis of such cases is of interest, primarily for the development of calculation methods, since the solution of the problem cannot be obtained in a closed form and can hardly be analyzed in details. We find a new approach to calculation of Mach surface of a point source rotating along a circle. Our method is based on Fourier transform and on further application of stationary phase approximation for the calculation of the coefficients of Fourier series. Besides, we discuss what features of the potentials can be analyzed in the framework of this approach. 

\section{Green's function and scalar potential for the radiation of rotating charge}
\label{S:2}
Let the point charge $q_9$ rotate along circle with a radius $R_0$ with constant angular frequency $\omega$. We use cylindrical coordinates $(r,\phi,z)$ to solve the wave equation for  Green’s function $G(r,z,\phi,t)$ or scalar potential $\Phi(r,z,\phi,t)$. Corresponding equations differ in the right-hand side. 
\begin{eqnarray}
\displaystyle&&\frac{1}{r}\frac{\partial}{\partial r}\left(r\frac{\partial G(r,z,\phi,t)}{\partial r}\right)+
\frac{\partial^2 G(r,z,\phi,t)}{\partial z^2}+\frac{1}{r^2}\frac{\partial^2 G(r,z,\phi,t)}{\partial \phi^2}-
\frac{1}{c^2}\frac{\partial^2 G(r,z,\phi,t)}{\partial t^2}\nonumber\\
&&=\rho_G (r,z,\phi,t),
\label{Eq1}
\end{eqnarray}
where
\begin{equation}
\displaystyle
\rho_G(r,z,\phi,t)=\frac{\delta(r-R_0)\delta(z-Z_0)\delta(\phi-\phi_0-\omega t)}{r}.
\label{Eq2}
\end{equation}
In (\ref{Eq2})  $(r,\phi,z)$ are the coordinates of the observation point and  
$\phi_0$ is the angle coordinate of the charge at $t=0$. The rotation circle lies in $z=Z_0$ plane with its center at $r=0$. 
The charge density in the right-hand side of wave equation for scalar potential is (CGS units)  
$\rho_{\Phi}(r,z,\phi,t)=-4\pi q_0\rho_G(r,z,\phi,t).$
Solution of  (\ref{Eq1}) ,  (\ref{Eq2})  can be represented as Lienard-Wiechert potential. This is a common way to investigate the radiation of superliminal charges ( see, e.g., \cite{BB}). 
To apply this method one should find the retarded time $t'$  as the solution of the equation 
\begin{equation}
\displaystyle
t=t'+\frac{1}{c}\sqrt{r^2+R_0^2-2rR_0\cos(\phi_0+\omega t')+(z-Z_0)^2}. 
\label{Eq3}
\end{equation}
This equation has more than one solution for superluminal charge $\omega R_0 > c$. The number of the solutions can be arbitrary large as the ratio $\omega R_0/c$ increases. An obvious drawback of this method is that the  (\ref{Eq3}) can not be solved analytically which hampers further analysis. 

Fourier transform is another way to construct the solution of (\ref{Eq1}) ,  (\ref{Eq2}) problem. This approach was mentioned in \cite{ArdavanPRD}. Let us assume that the solution of (\ref{Eq1}) ,  (\ref{Eq2}) depends on angle and time as the right hand side does, namely, $G(r,z,\phi,t)=G(r,z,\hat{\phi}=\phi-\phi_0-\omega t)$. Consistent application of Fourier transform in $\phi$ and $z$ results in the following representation of solution of  (\ref{Eq1}) ,  (\ref{Eq2}): 

\begin{equation}
\displaystyle
G(r,z,\hat{\phi})=\frac{1}{2\pi}\sum_{m=-\infty}^{\infty} G_m(r,z)\exp(i m \phi), 
 G_m(r,z)=\frac{1}{2\pi}\int_{-\infty}^{infty} \zeta_m(r,k)\exp(ikz)dk. 
\label{Eq4}
\end{equation}

Function $\zeta_m(r,k)$ satisfies the equation 
\begin{eqnarray}
\displaystyle&&
\frac{\partial^2\zeta_m(r,k)}{\partial r^2}+\frac{1}{r}\frac{\partial\zeta_m(r,k)}{\partial r}-\frac{m^2}{r^2}\zeta_m(r,k)+
\left(\frac{\omega^2 m^2}{c^2}-k^2\right)\zeta_m(r,k)\nonumber\\
&&=\exp(-ikZ_0)\frac{\delta(r-R_0)}{r}. 
\label{Eq5}
\end{eqnarray}
The solution of this equation depends on the sign of $\omega^2 m^2/c^2-k^2$. H. Ardavan called this type of equation 'mixed equation' \cite{ArdavanPRD}. Green's function does not have to be real. Still, the scalar potential should be a real-valued function. The potential can be obtained as a real part of Green’s function. The solution of nonhomogeneous equation (\ref{Eq5}) is not unique. We chose a specific solution aiming at simplification of further analysis. We apply Sommerfeld radiation condition for the indexes $m>0$:  

For $-\omega \vert m\vert/c<k<\omega \vert m\vert/c$, 

\begin{equation}
\displaystyle
\zeta_m(r,k)=\left[
\begin{array}{lr}
2i\pi^2 e^{-ikZ_0} \, H_m^{(1)}\left(\sqrt{\omega^2 m^2/c^2-k^2} R_0\right)\, J_m\left(\sqrt{\omega^2 m^2/c^2-k^2}\, r\right)\: r<R_0\\
2i\pi^2 e^{-ikZ_0} \, H_m^{(1)}\left(\sqrt{\omega^2 m^2/c^2-k^2}\, r\right)\, J_m\left(\sqrt{\omega^2 m^2/c^2-k^2} R_0\right)\: r>R_0\\
\end{array}
\right.
\label{Eq6}
\end{equation}

for $\vert k \vert >\omega \vert m\vert/c$, 

\begin{equation}
\displaystyle
\zeta_m(r,k)=\left[
\begin{array}{lr}
4\pi e^{-ikZ_0} \, K_m\left(\sqrt{\omega^2 m^2/c^2-k^2} R_0\right)\, I_m\left(\sqrt{\omega^2 m^2/c^2-k^2}\, r\right)\: r<R_0\\
4\pi e^{-ikZ_0} \, K_m\left(\sqrt{\omega^2 m^2/c^2-k^2}\, r\right)\, I_m\left(\sqrt{\omega^2 m^2/c^2-k^2} R_0\right)\: r>R_0\\
\end{array}
\right. , 
\label{Eq7}
\end{equation}
where $J_m, H_m^{(1)}, I_m, K_m$ are Bessel function of the first kind of order m, Hankel function of the first kind of order $m$, and modified Bessel function of the first and second kind of order m respectively. 

Therefore,
\begin{eqnarray}
\displaystyle&&
G_m(r,z)=\frac{1}{2\pi}\int_{-\infty}^{+\infty}\zeta_m(r,k)\exp(ikz)=\nonumber\\
&&=\frac{i}{2}\int\limits_{-\omega m/c}^{\omega m/c}e^{ik(z-Z_0)}H_m^{(1)}\left(\sqrt{\omega^2 m^2/c^2-k^2}\, r_{>}\right)\, J_m\left(\sqrt{\omega^2 m^2/c^2-k^2}\, r_{<}\right)dk+\nonumber\\
&&+\frac{1}{\pi}\int\limits_{\vert k \vert > \omega m/c} e^{ik(z-Z_0)}K_m\left(\sqrt{\omega^2 m^2/c^2-k^2}\, r_{>}\right)\, I_m\left(\sqrt{\omega^2 m^2/c^2-k^2}\, r_{<}\right)dk\nonumber\\
&&\equiv G_m^{(1)}(r,z)+G_m^{(2)}(r,z), 
\label{Eq8}
\end{eqnarray}
where $r_{<}=\textrm{min}\{r, R_0\}$, $r_{>}=\textrm{max}\{r, R_0\}$. For $m<0$ we replace Hankel function of the first kind by Hankel function of the second kind in the expression for $G_m^{(1)}(r,z)$. 

\section{Calculation of Fourier coefficients in stationary phase approximation}

We search for the equation of Mach surface as the condition of divergency of the series (\ref{Eq4}). To get this equation we need to calculate the integrals (\ref{Eq8}). For $z=Z_0$  some calculations can be performed analytically. We start with the application of 'multiplication theorem' \cite{Abramowitz} which can be derived from Graf’s addition theorem. For our purpose we use the following identities: 
\begin{eqnarray}
\displaystyle
H_m^{(1)}(S)J_m(s)=\frac{1}{2\pi}\int_{0}^{2\pi}H_0^{(1)}\left(\sqrt{S^2+s^2-2Ss\cos\theta}\right)exp(-im\theta)d\theta, \: S>s,\nonumber\\
K_m(S)I_m(s)=\frac{1}{2\pi}\int_{0}^{2\pi}K_0\left(\sqrt{S^2+s^2-2Ss\cos\theta}\right)exp(-im\theta)d\theta, \: S>s.\nonumber\\
\label{Eq9}
\end{eqnarray}
Applying (\ref{Eq9}) we write the first integral of (\ref{Eq8}) as follows: 
\begin{eqnarray}
\displaystyle
G_m^{(1)}(r,z)=\frac{i}{2\pi}\int_0^{2\pi} e^{-im\theta}d\theta \int_0^{\omega m/c}dk\,H_0^{(1)}\left(\sqrt{\omega^2 m^2/c^2-k^2}\sqrt{R_0^2+r^2-2R_0r\cos\theta}\right)
\label{Eq10}
\end{eqnarray}
We change the variables $\sqrt{\omega^2 m^2/c^2-k^2}=x$ and change the order of the integration. We substitute 
$H_0^{(1)}=J_0+iY_0,$ where $Y_0$ is Bessel function of second kind, into (\ref{Eq10}) and calculate the integrals with $J_0$ and $Y_0$ separately. 
The first integral is 
\begin{equation}
\displaystyle
\int\limits_0^{\omega m/c}\frac{J_0(R(\theta)x)x}{\sqrt{\omega^2 m^2/c^2-x^2}}dx=
\frac{\sin\left(m\omega R(\theta)/c\right)}{R(\theta)}, 
\label{Eq11}
\end{equation}
where we introuce the notation $R(\theta)=\sqrt{R_0^2+r^2-2R_0r\cos\theta}$. 
The second integral can be calculated as follows. 
For non-integer $\nu$ the following integral is known \cite{BrPrMr}:
\begin{equation}
\displaystyle
\int\limits_0^a\frac{x^{\nu+1}Y_{\nu}(cx)}{\sqrt{a^2-x^2}}dx=\frac{a^{\nu+1/2}}{sin{\nu\pi}}\sqrt{\frac{\pi}{2c}}
\left[\cos(\nu\pi)J_{\nu+1/2}(ac)-\mathbf{H}_{-\nu-1/2}(ac)\right]
\label{Eq12}
\end{equation}
where $\mathbf{H}$ is Struve function. This formula is correct for $a>0$ and $\Re[\nu]>-1$. 
To apply this formula for $\nu=0$ let us consider the limit $\nu \rightarrow 0$. Taking into account that 
$\mathbf{H}_{-1/2}(z)=J_{1/2}(z)$ we get 
\begin{eqnarray}
\displaystyle&&
\lim_{\nu\to 0}\frac{J_{\nu+1/2}(X)-\mathbf{H}_{-\nu-1/2}(X)}{\nu}=\nonumber\\
&&\lim_{\nu\to 0}\frac{J_{\nu+1/2}(X)-J_{1/2}(X)-\left(\mathbf{H}_{-\nu-1/2}(X)-(\mathbf{H}_{-1/2}(X)\right)}{\nu}=\nonumber\\
&&\left.\frac{\partial J_\mu (X)}{\partial \mu}\right|_{\mu=1/2}+\left.\frac{\partial \mathbf{H}_\mu (X)}{\partial \mu}\right|_{\mu=1/2}.\nonumber\\
\label{Eq13}
\end{eqnarray}
These derivatives were calculated in \cite{Geddes}, and, using these results, we get 
\begin{eqnarray}
\displaystyle&&
\int\limits_0^{\omega m/c}\frac{Y_0(R(\theta)x)x}{\sqrt{\omega^2 m^2/c^2-x^2}}dx=\nonumber\\
&&\frac{2}{\pi R(\theta)}\left(\sin(m\omega R(\theta)/c)\textrm{ci}(m\omega R(\theta)/c)-\cos(m\omega R(\theta)/c)\textrm{si}(m\omega R(\theta)/c)\right),\nonumber\\
\label{Eq14}
\end{eqnarray}
where $\textrm{ci}$ and $\textrm{si}$ are the integral cosine and sine. Therefore, we get $G_m^{(1)}(r,z=Z_0).$ 
The second integral over $k$ in (\ref{Eq8}) can also be calculated exactly:  
\begin{eqnarray}
\displaystyle&&
\int\limits_{\omega m/c}^{+\infty}K_0\left(\sqrt{k^2-\omega^2 m^2/c^2}\right)dk=\frac{1}{2R(\theta)}
\left[\pi \cos(R(\theta)\omega m/c)+\right.\nonumber\\
&&\left. 2\sin(R(\theta)\omega m/c)\textrm{ci}(R(\theta)\omega m/c)-2\cos(R(\theta)\omega m/c)\textrm{si}(R(\theta)\omega m/c)\right].\nonumber\\
\label{Eq15}
\end{eqnarray}
Combining (\ref{Eq11}), (\ref{Eq14}) and (\ref{Eq15}) we obtain
\begin{equation}
\displaystyle
G_m(r, z=Z_0)=\frac{1}{2\pi}\int\limits_0^{2\pi}\frac{e^{-im(\theta-\omega R(\theta)/c)}}{R(\theta)}d\theta=
\frac{1}{2\pi}\int\limits_0^{2\pi}\frac{e^{-im(\theta-\gamma\sqrt{1-2\rho\cos\theta+\rho^2})}}{R_0\sqrt{1-2\rho\cos\theta+\rho^2}}, 
\label{Eq16}
\end{equation}
where we introduce the notations $\rho=r/R_0$, $\gamma=\omega R_0/c$. Though contribution (\ref{Eq15}) into 
$G_m(r,z)$ cannot lead to the divergency of the series (\ref{Eq4}), and only the contribution of $G_m^{(1)}(r,z=Z_0)$ is important, we take 
the sum $G_m^{(1)}(r, z=Z_0)+G_m^{(2)}(r, z=Z_0)$ to get a simple form (\ref{Eq16}). 

The divergency of series (\ref{Eq4}) is determined by the divergency of its remainder starting from arbitrary large $m$. Therefore, to get the divergency condition 
we can take index $m$ as large as we need for the estimations of the series' terms. In particular, we can apply stationary phase approximation to estimate (\ref{Eq16}). 
The stationary point is determined by zero of the exponent and we easily find the equation for the stationary point $\theta_0$:
\begin{equation}
\displaystyle
1-\frac{\gamma\rho\sin\theta_0}{\sqrt{1-2\rho\cos\theta_0+\rho^2}}=0
\label{Eq17}
\end{equation}
Taking into account the inequalities $\gamma>1$, $\rho>1/\gamma$ we find 
\begin{equation}
\displaystyle
\theta_{0\pm}=\arccos\frac{1\pm \sqrt{(\gamma^2-1)(\gamma^2\rho^2-1)}}{\gamma^2\rho}. 
\label{Eq18}
\end{equation}
Using (\ref{Eq18}) we calculate $G_m(r,z)$ in stationary phase approximation. 
We get the following approximate expression for the remainder of series (\ref{Eq4})
\begin{equation}
\displaystyle
\sum_{m>M_0}\sum_{\sigma=\pm}\frac{\exp\left(im\left(\hat{\phi}-\theta_{0\pm}+\gamma\sqrt{1-(2/\gamma^2)\left(1+\sigma\sqrt{(\gamma^2-1)(\gamma^2\rho^2-1)}\right)}\right)\right)}{\sqrt{m\left(1-(2/\gamma^2)\left(1+\sigma\sqrt{(\gamma^2-1)(\gamma^2\rho^2-1)}\right)\right)}}, 
\label{Eq19}
\end{equation}
where $M_0$ is an index from which we start the remainder of series (\ref{Eq4}). 
We choose $\sigma=+1$ for $\theta_{0+}$ and $\sigma=-1$ for $\theta_{0-}$ in (\ref{Eq19}).
As we see, the remainder of series (\ref{Eq4}) calculated within stationary phase approximation 
is nothing but the remaider of series for polylogarithm, $\textrm{Li}_{1/2}(x)=\sum_{k=1}^{\infty}x^k/\sqrt{k}$. Polylogarithm $\textrm{Li}_{1/2}$ turns into infinity 
for $x=1$. Consequently, the condition of the divergency of (\ref{Eq19}) is following:  
\begin{equation}
\displaystyle
\hat{\phi}_{\pm}=\theta_{\pm}-\frac{\omega}{c}\sqrt{R_0^2+r^2-2R_0r\cos\theta_{\pm}}. 
\label{Eq20}
\end{equation}  
This equation together with (\ref{Eq19}) determines $\hat{\phi}$ as the function of $r$. The corresponding equation in polar coordinates is following 
\begin{equation}
\displaystyle
\left\{
\begin{array}{ll}
x_{\pm}=r\cos\phi_{\pm}\\
y_{\pm}=r\cos\phi_{\pm}
\end{array} \right.
\label{Eq21}
\end{equation}
Fig. 1 displays the lines determined by the equations (\ref{Eq21}) for $c=5$, $\omega=3.5$, $R_0=3.5$, the blue line represents the branch marked by index ”+”, whereas the pink line represents the branch marked by index “-”.  

Therefore, we get the equation of the cross-section of Mach surface with the rotation plane of the point source. 
The equation for the Mach surface can be derived as the equation of the envelope (e.g., see (\cite{BB})). Depending on the parametrization, this equation can be of diferent form. The obtained representaion Eqs.(\ref{Eq21}) differs from the one obtained in (\cite{BB}) and similar to the representation derived as an envelope function in (\cite{ArdavanPRE}). 

\begin{figure}[tp]
\includegraphics[width=85mm,scale=0.85]{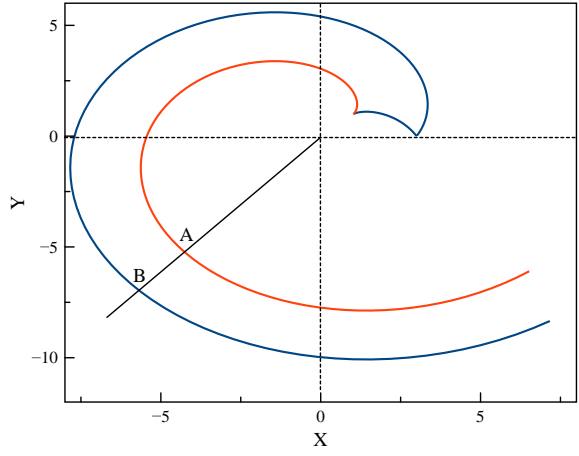}
\caption{\label{Figure1} The cross-section of Mach cone with $z=Z_0$ plane for $c=5$, $\omega=3.5$, $R_0=3.5$. The diameter of Mach cone 
$D\sim |AB|.$}
\end{figure}

In general case $z\ne Z_0$ the calcultions are more cumbersome. We need to use stationary phase approximation two times. Like in the case $z=Z_0$ 
we start with the multiplication  theorem, and for the first integral $G_m^{(1)}(r,z)$ we apply the change of the variables 
\begin{equation}
\displaystyle
k=-\frac{\omega m}{c}\sqrt{1-s^2}, k\in [-\omega m/c, 0]; k=\frac{\omega m}{c}\sqrt{1-s^2}, k\in [0, \omega m/c]. 
\label{Eq22}
\end{equation}
Therefore, we get
\begin{eqnarray}
\displaystyle&&
G_m^{(1)}(r,z)=\frac{i}{4\pi}\int\limits_0^{2\pi}\int\limits_0^{1} e^{i(m\omega/c)\sqrt{1-s^2}\,(z-Z_0)} 
H_0^{(1)}\left(\frac{m\omega}{c}R(\theta)s\right)
\frac{(m\omega/c))sds}{\sqrt{1-s^2}} e^{-im\theta}d\theta+\nonumber\\
&&\frac{i}{4\pi}\int\limits_0^{2\pi}\int\limits_0^{1} e^{-i(m\omega/c)\sqrt{1-s^2}\,(z-Z_0)} 
H_0^{(1)}\left(\frac{m\omega}{c}R(\theta)s\right)
\frac{(m\omega/c))sds}{\sqrt{1-s^2}}e^{-im\theta}d\theta .\nonumber\\
\label{Eq23}
\end{eqnarray}
We start with the integrals over variable $s$. 
Let us chose function $\delta(m)$ such that $0<\delta(m)\ll <1$ and $\lim_{m\to \infty}\delta(m)=0$, 
$\lim_{m\to\infty}m\delta(m)=\infty$,  $\lim_{m\to\infty}m\delta(m)^2=0$. For instance, we can choose $\delta(m)=m^{-\alpha}$, $1/2<\alpha<1$. We divide the integration interval into two sub-intervals, $[0, \delta(m)]$ and $[\delta(m), 1]$. The integral over the first one is calculated as follows.  

\begin{eqnarray}
\displaystyle&&
\int\limits_0^{m^{-\alpha}} e^{i\frac{m\omega}{c}\sqrt{1-s^2}\,(z-Z_0)-im\theta}
H_0^{(1)}\left(\frac{m\omega}{c}R(\theta)s\right)\frac{(m\omega/c)sds}{\sqrt{1-s^2}}\approx\nonumber\\
&&\int\limits_0^{m^{-\alpha}} e^{i\frac{m\omega}{c}\,(z-Z_0)-im\theta}
H_0^{(1)}\left(\frac{m\omega}{c}R(\theta)s\right)(m\omega/c)sds\approx\nonumber\\
&&e^{i\frac{m\omega}{c}(z-Z_0)-im\theta}\left\{\frac{2i}{\pi R(\theta)^2 (m\omega/c)}+H_1^{(1)}\left((m^{1-\alpha}\omega/c)R(\theta)\right)\right\}\approx\nonumber\\
&&e^{i\frac{m\omega}{c}(z-Z_0)-im\theta}\left\{\frac{2i}{\pi R(\theta)^2 (m\omega/c)}+\sqrt{\frac{2m^{-\alpha}}{\pi R(\theta)(m\omega/c)}}e^{i((m^{1-\alpha}\omega/c)R(\theta)-3\pi/4)}\right\}.\nonumber\\
\label{Eq24}
\end{eqnarray}
Similar expression is obtained for the second integral of (\ref{Eq23}). 
The well-known expression for the asymptotic expression for Hankel function of the first kind, $H_m^{(1)}(X)\sim \sqrt{2/(\pi X)}\exp(i(X-m\pi/2-\pi/4))$ 
was used in (\ref{Eq24}). This asymptotic form is valid for $X\gg m^2$. Due to application of multiplication theorem now we deal with Hankel function with index $m=0$ and such condition is true for any positive argument of Hankel function. Using this asymptotic expresion, we estimate the integral over $[m^{-\alpha}, 1]$ in stationary phase approximation. For 
$s \in [m^{-\alpha}, 1]$  the argument of Hankel function in the integrand can be made arbitrarily large by choosing a large enough $m$. 
Replacing Hankel function by its asymptotic expression, we get  
\begin{eqnarray}
\displaystyle&&
\int\limits_{m^{-\alpha}}^1 e^{i(m\omega/c)\sqrt{1-s^2}\,(z-Z_0)-im\theta}
H_0^{(1)}\left(\frac{m\omega}{c}R(\theta)s\right)\frac{(m\omega/c)sds}{\sqrt{1-s^2}}+\nonumber\\
&&\int\limits_{m^{-\alpha}}^1 e^{-i(m\omega/c)\sqrt{1-s^2}\,(z-Z_0)-im\theta}
H_0^{(1)}\left(\frac{m\omega}{c}R(\theta)s\right)\frac{(m\omega/c)sds}{\sqrt{1-s^2}}\nonumber\\
\label{Eq25}
\end{eqnarray}
Integrals (\ref{Eq25}) can be calculated in stationary phase approximation. Depending on sign of  $z-Z_0$, 
only one integral of the two gives the contribution which can lead to the divergency of series (\ref{Eq4}). 
It is the first integral for $z > Z_0$. It is calculated in stationary phase approximation as follows:   
\begin{eqnarray}
\displaystyle&&
\frac{m\omega}{c}\int\limits_{m^{-\alpha}}^1 e^{i(m\omega/c)\sqrt{1-s^2}\,(z-Z_0)-im\theta}
H_0^{(1)}\left(\frac{m\omega}{c}R(\theta)s\right)\frac{(m\omega/c)sds}{\sqrt{1-s^2}}\approx\nonumber\\
&&\frac{m\omega}{c}\int\limits_{m^{-\alpha}}^1
\frac{\sqrt{2}e^{i\frac{m\omega}{c}\left(\sqrt{1-s^2}(z-Z_0)+R(\theta)s\right)-i\pi/4-im\theta}}{\sqrt{\pi\frac{m\omega}{c}R(\theta)s}}\frac{s\,ds}{\sqrt{1-s^2}}
\approx\nonumber\\
&&\frac{e^{i\left((m\omega/c)\sqrt{R_0^2+r^2-2R_0 r \cos\theta+(z-Z_0)^2}-m\theta\right)}}
{\sqrt{R_0^2+r^2-2R_0 r \cos\theta+(z-Z_0)^2}}
\label{Eq26}
\end{eqnarray}


For $z < Z_0$ the second integral of (\ref{Eq25}) gives the same contribution. As a result, the sum (\ref{Eq25}) is the same expression as (\ref{Eq16}) with the replacement 
$R(\theta) \rightarrow \sqrt{R_0^2+r^2-2R_0 r \cos\theta+(z-Z_0)^2}$. 
We calculate the integral 
\begin{equation}
\displaystyle
\int\limits_0^{2\pi}
\frac{e^{i\left((m\omega/c)\sqrt{R_0^2+r^2-2R_0 r \cos\theta+(z-Z_0)^2}-m\theta\right)}} d\theta
\label{Eq27}
\end{equation}
in stationary phase approximation as it was made for $z=Z_0$. Correspondingly, Max surface is determined by the relations,  
\begin{eqnarray}
\displaystyle&&
\hat{\phi}_{\pm}=\arccos\frac{1\pm\sqrt{1-\gamma^2-\rho^2 \gamma^2-\gamma^2 \xi^2+\gamma^4 \rho^2}}{\gamma^2 \rho}\nonumber\\
&&-\gamma\sqrt{1+\rho^2-2\rho\cos\theta+\xi^2},
\label{Eq28}
\end{eqnarray}
where $\xi=(z-Z_0)/R_0$. Eq.(\ref{Eq28}) is similar to formula (20). Equation (\ref{Eq28}) together with (\ref{Eq21}) gives the cross-section of Mach surface with arbitrary $z=const$  plane. 
The analysis of the part of the series (\ref{Eq4}) with $m<0$  leads to the same equations for Mach surface. 

\section{Max surface characteristics in the large $\omega$  limit}

Thus, we presented the derivation of Mach surface using stationary phase approximation for Green’s function (or scalar potential) of superluminal point source rotating along a circle. This approach is convenient for obtaining some characteristics of the potentials and field. One of the examples of such analysis is the asymptotic dependence of Green's function (or scalar potential) on the rotation frequency $\omega$ for large $\omega$. 

We note, that $\omega$ appears as a multiplier in the exponent of (\ref{Eq28}). Therefore, stationary phase approximation is applicable for any $m\ne 0$  if $\omega$ is large enough. So, we find that $G(r,z,\phi)$ decreases as $1/\sqrt{omega}$  for large  $\omega$. 
  
The other example is the estimation of the diameter of cross-section of the Mach surface for large $\gamma=\omega R_0/c$ and $rho=r/R_0$ . In the case of rectilinear motion, the diameter $D$ of cross-section of Mach cone increases linearly 
as the function of distance $S$  from the source: $D=2S/\sqrt{(u/c)^2-1}$, where $u$ is the velocity of the point charge. In the case of point charge rotating along a circle, the cross-section size of Mach surface 
can be estimated as the length of a segment $[A, B]$ (see Fig. 1) of a ray 
beginning at the origin and contained inside Mach surface (segment $[A, B]$ in Fig. 1). We get the following approximate expressions from the Eqs. (\ref{Eq19}), (\ref{Eq20}) : 
\begin{equation}
\displaystyle
\hat{\phi}_{+}\approx -\gamma (\rho-1),\: \; \hat{\phi}_{-}\approx-\pi-\gamma(\rho+1), \: \gamma\gg 1, \rho\gg 1
\label{Eq29}
\end{equation}
Let us define $\beta$ by the following euality: 
\begin{equation}
\displaystyle
\gamma\approx=\pi N+\beta, \; \; 0<\beta<\pi
\label{Eq30}
\end{equation}
Then,
\begin{equation}
\displaystyle
\hat{\phi}_{+}\approx \pi N+\beta-\gamma\rho, \: \hat{\phi}_{-}\approx -\pi N-\beta+\pi-\gamma\rho
\label{Eq31}
\end{equation}
In general case, $\hat{\phi}_{+}$ and $\hat{\phi}_{-}$ are unequal and for the same $\rho$ the points $(x_+, y_+)$ and $(x_-, y_-)$ belong to different rays. To  belong to the same ray, $\hat{\phi}_+$ and $\hat{\phi}_-$ must be defined for different $\rho$.  
Let us take $\rho$ for $\hat{\phi}_+$ and $\rho-\Delta\rho$ for $\hat{\phi}_-$, i.e. 
\begin{equation}
\displaystyle
\phi_{-}^{(1)}\approx -\pi N-\beta+\pi-\gamma(\rho-\Delta\rho), \: \Delta\rho>0, 
\label{Eq32}
\end{equation}
$N$ is integer. Now, to get segment $[A, B]$ contained inside the Max surface (Fig. 1) let us consider difference 
\begin{equation}
\displaystyle
\hat{\phi}_{+} - \hat{\phi}_{-}^{(1)}\approx 2\beta-\pi-\gamma\Delta\rho+2\pi N, 
\label{Eq33}
\end{equation}
The minimal $\Delta\rho>0$, such that $\hat{\phi}_{+}=\hat{\phi}_{-}[\textrm{mod}\;2\pi]$, is
\begin{equation}
\displaystyle
\Delta\rho=\left[
\begin{array}{lr}
(2\beta-\pi)/\gamma, 2\beta-\pi>0\\
(2\beta+\pi)/\gamma, 2\beta-\pi<0\\
\end{array}
\right.
\label{Eq34}
\end{equation}
So, the diameter of the curved Mach cone for large distance from origin and high rotation frequency is 
\begin{equation}
D\approx R_0\Delta\rho=(c/\omega)(2\beta\pm\pi)
\label{Eq35}
\end{equation}
We see, that the diameter of the curved Mach cone does not increase as we move away from the origin and that it is proportional to  $\omega^{-1}$. 

\section{Conclusion}

In conclusion, we propose a new way to get Mach surface for the electromagnetic radiation of a point source rotating along a circle with constant speed. In the framework of proposed approach, we investigate some characteristics of this Mach surface for large frequency rotation of the source. Though there are no point sources (even fictitious ones) which move superluminally in vacuum, the presented analysis, besides its model aspect, also has some practical value, since the constructed scalar potential is also a Green’s function of the system.

\textbf{Acknowledgements}.
M.Ye. Zhuravlev is grateful to Russian Foundation for Basic Research for the support through the Project No 19-02-00316-A.

\section*{References}

\bibliography{StPhMachA}

\end{document}